\documentclass[prl,twocolumn,showpacs]{revtex4}
\usepackage{textcomp}
\usepackage{amsmath}
\usepackage{graphicx}

\begin{document}
\title{Submicrometer position control of single trapped neutral atoms }
    \author{I.~Dotsenko}
    \author{W.~Alt}
    \author{M.~Khudaverdyan}
    \author{S.~Kuhr}
    \author{D.~Meschede}
    \author{Y.~Miroshnychenko}
    \author{D.~Schrader}
    \author{A.~Rauschenbeutel}
    \email{rauschenbeutel@iap.uni-bonn.de}
    \affiliation{Institut f\"ur Angewandte Physik, Universit\"at Bonn,
    Wegelerstra\ss e 8, D-53115 Bonn, Germany}
\date{\today}
\pacs{32.80.Lg, 32.80.Pj, 39.25.+k, 42.30.-d}

\begin{abstract}

We optically detect the positions of single neutral cesium atoms
stored in a standing wave dipole trap with a sub-wavelength
resolution of 143~nm rms. The distance between two simultaneously
trapped atoms is measured with an even higher precision of 36~nm
rms. We resolve the discreteness of the interatomic distances due
to the 532~nm spatial period of the standing wave potential and
infer the exact number of trapping potential wells separating the
atoms. Finally, combining an initial position detection with a
controlled transport, we place single atoms at a predetermined
position along the trap axis to within 300~nm rms.

\end{abstract}

\maketitle

Precision position measurement and localization of atoms is of
great interest for numerous applications and has been achieved in
and on solids using e.g.~scanning tunnelling microscopy
\cite{Binnig99}, atomic force microscopy \cite{Giessibl03}, or
electron energy-loss spectroscopy imaging \cite{Suenaga00}.
However, if the application requires long coherence times, as is
the case in quantum information processing \cite{DiVincenzo95 and
Ekert96} or for frequency standards, the atoms should be well
isolated from their environment. This situation is realized for
ions in ion traps, freely moving neutral atoms, or neutral atoms
trapped in optical dipole traps. For the case of ions, positions
\cite{Walther01, Blatt02} and distances \cite{Wineland03} have
been optically measured and controlled with a sub-optical
wavelength precision. Similar precision has been reached in an
all-optical position measurement of freely moving atoms
\cite{Thomas95}. Dipole traps, operated as optical tweezers, have
been used to precisely control the position of individual neutral
atoms \cite{Kuhr01,Bergamini04}. To our knowledge, however, a
sub-micrometer position or distance measurement has so far not
been achieved in this case. Such a control of the relative and
absolute position of single trapped neutral atoms, however, is an
important prerequisite for cavity quantum electrodynamics as well
as cold collision experiments, aiming at the realization of
quantum logic operations with neutral atoms.

Here, we report on the measurement and control of the position of
single neutral atoms stored in a standing wave optical dipole trap
(DT). The positions of the atoms are inferred from their
fluorescence using high resolution imaging optics in combination
with an intensified CCD camera (ICCD). The absolute position of
individual atoms along the DT is measured with a precision of
143~nm rms. The relative position of the atoms, i.e.~their
separation, is determined more accurately by averaging over many
measurements, yielding a relative position uncertainty of 36~nm.
Due to this high resolution, we can resolve the discreteness of
the distribution of interatomic distances in the standing wave
potential even though our DT is formed by a Nd:YAG laser with
potential wells separated by only 532~nm. This allows us to
determine the exact number of potential wells between
simultaneously trapped atoms. Finally, using our ``optical
conveyor belt'' technique \cite{Kuhr01,Schrader01} we transport
individual atoms to a predetermined position along the DT axis
with an accuracy of 300~nm, thereby demonstrating a high degree of
control of the absolute atom position.

\begin{figure}[b]
    \begin{center}
        \includegraphics[width=68mm]{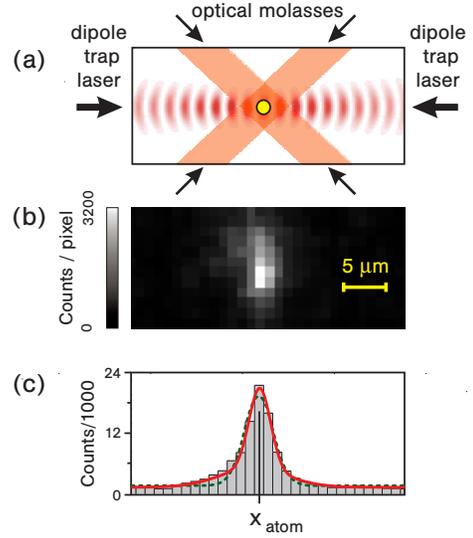}
    \end{center}
\caption{Determination of the position of a single trapped atom.
(a) An atom stored in the standing wave dipole trap is illuminated
with an optical molasses (schematic drawing). (b) ICCD image of
one atom stored in the DT with an exposure time of 1~s. The
observed fluorescence spot corresponds to about 200 detected
photons. (c) To determine the position of the atom along the DT
axis, the pixel counts are binned in the vertical direction. The
solid line corresponds to the line spread function of our imaging
optics and reveals the absolute position $x_\mathrm{atom}$ of the
atom. For comparison, the dashed line shows a fit with a simple
Gaussian (see text for details).}
    \label{fig:PositionDetermination}
\end{figure}

We will only give the essential details of our experimental
set-up. A more exhaustive description can e.g.~be found in
\cite{Schrader01, Miroshnychenko03}. Our standing wave dipole trap
is formed by the interference pattern of two counter-propagating
Nd:YAG laser beams ($\lambda =1.064\, \mu$m) with a waist of
$2w_0=38\,\mu$m and a total optical power of $2$~W. They produce a
trapping potential with a maximal depth of
$U_\mathrm{max}/k_\mathrm{B}=0.8$~mK for cesium atoms. Mutually
detuning the frequency of the two laser beams moves the trapping
potential along the DT axis and thereby transports the atoms
\cite{Kuhr01, Schrader01, Miroshnychenko03}. The laser frequency
is changed by acousto-optic modulators (AOMs), placed in each beam
and driven by a digital dual-frequency synthesizer. The DT is
loaded with cold atoms from a high-gradient magneto-optical trap
(MOT). We deduce the exact number of atoms in the MOT from their
discrete fluorescence levels detected by an avalanche photodiode.
The transfer efficiency between the traps is better than 99~\%.

In order to obtain fluorescence images of the atoms in the DT, we
illuminate them with a near-resonant three-dimensional optical
molasses, see Fig.~\ref{fig:PositionDetermination}(a). The
molasses counteracts heating through photon scattering, resulting
in an atom temperature of about 70~$\mu$K. The storage time in the
trap is about $25$~s, limited by background gas collisions. The
fluorescence light is collected by a diffraction limited
microscope objective \cite{Alt02} and imaged onto the ICCD
\cite{Miroshnychenko03}. One detected photon (quantum efficiency
approx.~10~\% @ 852~nm) generates on average 350~counts on the CCD
chip, and one $13\ \mu\mathrm{m}\times 13\ \mu\mathrm{m}$ CCD
pixel corresponds to $0.933\,(\pm0.004)\,\mu$m in the object
plane.

Figure~\ref{fig:PositionDetermination}(b) shows an ICCD image of a
single atom stored in the DT with an exposure time of 1~s. This
exposure time is much longer than the timescale of the thermal
position fluctuations of the atom inside the trap. Therefore, the
vertical width of the fluorescence spot, i.e.~perpendicular to the
DT axis, is essentially defined by the spread of the Gaussian
thermal wave packet of the atom in the radial direction of the
trap. In the axial direction of the DT, the wave packet has a much
smaller $1/\sqrt{e}$-halfwidth of only $\Delta
x_\mathrm{therm}=35$--50~nm, depending on the depth of the DT. In
addition to these thermal fluctuations, the axial position of the
standing wave itself is fluctuating by $\sigma_\mathrm{fluct}(1\,
\mathrm{s})= 42\,(\pm 13)$~nm during the 1~s exposure time due to
drifts and acoustic vibrations of the optical setup (see below).
The horizontal $1/\sqrt{e}$-halfwidth of the detected fluorescence
peak, $w_\mathrm{ax}=1.3(\pm$0.15)~$\mu$m, is much larger and is
caused by diffraction within the imaging optics and a slight
blurring in the intensification process of the ICCD. Compared to
the point spread function of our imaging system, $\Delta
x_\mathrm{therm}$ and $\sigma_\mathrm{fluct}$ have thus a
negligible effect on $w_\mathrm{ax}$. Note, however, that all the
atom positions and the distances between atoms given in the
following refer to the {\em center} of the Gaussian thermal wave
packets of the atoms.

The ICCD image is characterized by its intensity distribution
$I(\tilde{x}_i,\tilde{y}_j)$, where $\tilde{x}_i$ and
$\tilde{y}_j$ denote the horizontal and vertical position of pixel
$\{i,j\}$, respectively. In order to determine the horizontal
position $\tilde{x}_\mathrm{atom}$ of the fluorescence peak from
the ICCD image, we bin $I(\tilde{x}_i,\tilde{y}_j)$ in the
vertical direction. Neglecting noise for the moment, this yields
$I(\tilde{x}_i)=\sum_{j} I(\tilde{x}_i,\tilde{y}_j)\propto
L(\tilde{x}_i-\tilde{x}_\mathrm{atom})$, where $L(\tilde{x})$ is
the line spread function (LSF) of our imaging optics. Without
distortions, the object coordinate $x_\mathrm{atom}$ and the image
coordinate $\tilde{x}_\mathrm{atom}$ are connected by the relation
$x_\mathrm{atom}=(\tilde{x}_\mathrm{atom}-
\widetilde{\mathcal{O}}_x)/M$, where $\widetilde{\mathcal{O}}_x$
is the image coordinate of the origin and $M$ is the magnification
of our imaging optics. In general, $\widetilde{\mathcal{O}}_x$ and
$M$ have to be calibrated from independent measurements. In the
present case, however, no physical point in space is singled out
as an origin and we arbitrarily set
$\widetilde{\mathcal{O}}_x\equiv 0$.

Our LSF is position-independent and is well described by a sum of
two Gaussians with a ratio of 4.4:1 in heights and 1:3.2 in
widths, with a slight horizontal offset with respect to each
other, see Fig.~\ref{fig:PositionDetermination}(c). We define
$\tilde{x}_\mathrm{atom}$ as the position of the maximum of this
LSF. In our experiment, it is determined by fitting a simple
Gaussian to the fluorescence peak. This procedure has been chosen
because it can be carried out in a fast automated way, yielding
information about the atom position during the running
experimental sequence. Assuming pure shot noise, this allows to
determine $x_\mathrm{atom}$ with a statistical error of
    \begin{equation}\label{eq:deltaX}
        \Delta x_\mathrm{stat}= 1.44\,
        w_\mathrm{ax}/\sqrt{N_\mathrm{ph}},
    \end{equation}
where $N_\mathrm{ph}$ is the number of detected photons and the
numerical factor has been determined by a numerical simulation
taking into account the experimental LSF and the bin size. In the
experiment the value of $N_\mathrm{ph}$ depends on the
illumination parameters. Here, $N_\mathrm{ph}=200(\pm 30)$ photons
per second per atom, so that $\Delta x_\mathrm{stat}= 130\,
(\pm20)$~nm. Our simulation also yields a constant position offset
of 42~nm of the fitted center of the Gaussian with respect to the
maximum of the LSF, due to the slight asymmetry of our LSF. This
offset only leads to a global shift of $\widetilde{\mathcal{O}}_x$
and is irrelevant for our analysis.

In addition to the statistical error, two further sources
influence the precision of the position detection: the background
noise of our ICCD image and the position fluctuations of the DT.
The background in Fig.~\ref{fig:PositionDetermination} originates
in equal proportions from stray light and the read-out process of
the ICCD, yielding a total offset of $2300(\pm300)$ counts per bin
for 1~s exposure time. The noise of $300$ counts per bin
introduces an additional uncertainty of $\Delta x_\mathrm{backgr}
= 15$~nm to the fitted peak center.

The atom position is subject to position fluctuations of the DT,
$\sigma_\mathrm{fluct}$. Since $\sigma_\mathrm{fluct}$ cannot be
extracted from the ICCD image, we determine it in an independent
measurement. For this purpose, we mutually detune the two trap
beams and overlap them on a fast photodiode. From the phase of the
resulting beat note we infer the phase variations $\phi(t)$ of the
standing wave with a 300~kHz bandwidth. The standard deviation of
$\phi(t)$, $\sigma_\phi(\tau)$, is directly related to the
position fluctuations of the DT during the time interval $\tau$ by
$\sigma_\mathrm{fluct}(\tau) = \lambda/2
\cdot\sigma_{\phi}(\tau)/2\pi$. We have found
$\sigma_\mathrm{fluct}(1\,\mathrm{s}) =42\,(\pm 13)$~nm. Thus,
using the approximation of Gaussian-distributed position
fluctuations, which we have checked to be valid to better than
1~\% in our case, the position uncertainty immediately after the
1~s exposure time is given by
    \begin{equation}\label{eq:deltaX_a}
        \Delta x_\mathrm{atom}^2(1\,\mathrm{s})=
        \Delta x_\mathrm{stat}^2+
        \Delta x_\mathrm{backgr}^2 +
        \sigma_\mathrm{fluct}^2(1\,\mathrm{s})\ ,
    \end{equation}
yielding $\Delta x_\mathrm{atom}(1\,\mathrm{s})= 140\,(\pm
20)$~nm.

Finally, the read-out and the data analysis of the image take an
additional 0.5~s during which $x_\mathrm{atom}$ is further subject
to position fluctuations of the DT. This increases the variance of
the position measurement by
$2\sigma_\mathrm{fluct}^2(0.5\,\mathrm{s})$. Thus, we can
determine the absolute position of the trapped atom with a
precision of  $\Delta x_\mathrm{atom}(1.5\,\mathrm{s})= 143\,(\pm
20)$~nm within 1.5~s (1~s exposure time plus 0.5~s read-out and
data analysis). Our analysis shows that this precision cannot be
significantly increased by extending the exposure time because the
benefit of the higher photon statistics for longer times is
counteracted by the increase in $\sigma_\mathrm{fluct}(\tau)$.

\begin{figure}[b!]
    \begin{center}
        \includegraphics[width=66mm]{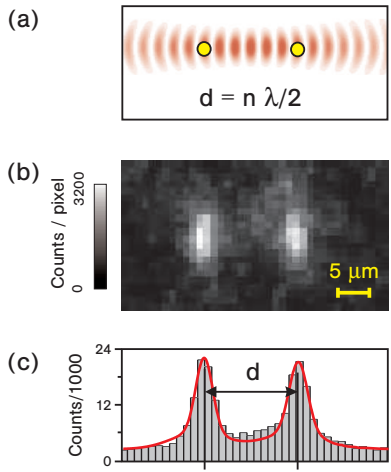}
    \end{center}
\caption{Determination of the distance between atoms. (a) Two
atoms in the standing wave dipole trap have a separation of $n
\lambda/2$ with $n$ integer. (b) After loading the atoms into the
DT, we successively take many camera pictures of the same pair of
atoms. (c) From each picture (exposure time 1~s) we determine the
positions of the atoms and their separation $d$. Averaging over
many measurements of $d$ reduces the statistical error and allows
us to infer $n$.}
    \label{fig:AtomSeparationScheme}
\end{figure}

While for some applications the absolute position of the atoms
must be known to the highest possible precision, other
experiments, like e.g.~controlled cold collisions \cite{Mandel03},
require a precise knowledge of the separation $d$ between atoms.
In the following we will show that in our case this separation can
be more precisely determined than the absolute positions of the
individual atoms. The reason is that DT fluctuations equally
influence all simultaneously trapped atoms and therefore do not
affect the separation between them. Thus, this distance can be
averaged over many measurements. Given the precision of the peak
detection, the uncertainty of the separation $d$ between two atoms
determined from {\em one} picture should be $\Delta d^2=2(\Delta
x_\mathrm{stat}^2+\Delta x_\mathrm{backgr}^2)$. Averaging the
results from $N_\mathrm{pic}$ images should then reduce the
statistical error of the mean value $\bar{d}$ to $\Delta
\bar{d}=\Delta d/\sqrt{N_\mathrm{pic}}$. Since the data processing
in this case is carried out at a later stage, we use the
experimentally established LSF [see Fig.~1(c)] for fitting the
fluorescence peaks. For the case of partially overlapping
fluorescence spots $(d \lesssim 10~\mu\mathrm{m})$, this method
yields more precise results for the two atom positions than
fitting a simple Gaussian. For $d\lesssim 4~\mu\mathrm{m}$ the
increasing overlap reduces the precision of the position
determination. We have therefore restricted our investigations to
the case where the atoms are separated by more than 4~$\mu$m.

To realize this scheme experimentally, we first load the DT with
two atoms, see Fig.~\ref{fig:AtomSeparationScheme}(a). We then
typically take $N_\mathrm{pic} =10$ successive camera pictures of
the same pair before one of the two atoms leaves the trap. In
these experiments, we detect $N_{\mathrm{ph}}=270(\pm 30)$ photons
per second per atom. From each picture we determine the distance
$d$ between the atoms, see Fig.~\ref{fig:AtomSeparationScheme}~(b)
and (c), and then calculate its mean value $\bar{d}$ and its
standard deviation $\Delta d$ for each pair. Our measured value of
$\Delta d =135\,(\pm30)$~nm is in resonable agreement with the
expected value of $\sqrt{2}\Delta x_\mathrm{stat}=160(\pm 25)$~nm,
inferred from Eq.~(\ref{eq:deltaX}), thereby confirming its
validity.

\begin{figure}[b!]
    \begin{center}
        \includegraphics[width=72mm]{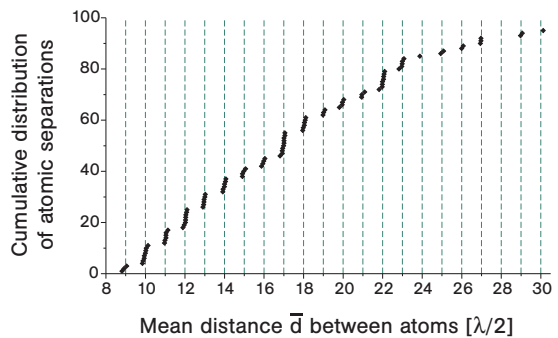}
    \end{center}
\caption{Cumulative distribution of separations between atoms in
the dipole trap measured with the scheme presented in
Fig.~\ref{fig:AtomSeparationScheme}. In order to resolve the
periodic structure of the trap, we reduce the statistical error in
the measurement of the atomic separation by averaging over more
than 10 distance measurements for each atom pair. The
discretization of the distances to $n\lambda/2$ is clearly visible
in the data. Our resolution is thus sufficient to determine the
exact number of potential wells between any two optically resolved
atoms.}
    \label{fig:DiskreteDistances}
\end{figure}

By averaging the distance over about $N_\mathrm{pic}=10$ images
per atom pair, the uncertainty in $\bar{d}$ should therefore be
reduced to $\Delta \bar{d}\approx 40$~nm. Now, the separation of
trapped atoms equals $d= n\lambda/2$ with $n$ integer, see
Fig.~\ref{fig:AtomSeparationScheme}(a). If $d$ can be measured
with a precision $\Delta d\ll\lambda/2$, its distribution should
therefore reveal the standing wave structure of the DT. Indeed,
the $\lambda/2$ period is strikingly apparent in
Fig.~\ref{fig:DiskreteDistances} which shows the cumulative
distribution of mean distances $\bar{d}$ between atoms.
The resolution of our distance measurements can be directly
inferred from the finite width of the steps observed in
Fig.~\ref{fig:DiskreteDistances}, yielding $\Delta
\bar{d}=36\,(\pm 12)$~nm. This result proves that we can determine
the exact number of potential wells between two optically resolved
atoms, a situation that so far seemingly required much longer
(e.g.~CO$_2$) trapping laser wave lengths \cite{Scheunemann00}.

Using our scheme to precisely measure the position of an atom, we
now demonstrate active control of its absolute position along the
DT axis. This is realized by transporting the atom to a
predetermined position $x_\mathrm{target}$ by means of our optical
conveyor belt. Initially, we determine the position of the atom
and its distance $L$ from $x_\mathrm{target}$ from an ICCD image
by fitting a simple Gaussian. To move the atom to
$x_\mathrm{target}$, it is uniformly accelerated along the first
half of $L$ and uniformly decelerated along the second half with
an acceleration of $a=\pm 1000$~m/s$^2$ \cite{Schrader01}. To
confirm the successful transport to $x_\mathrm{target}$, we take a
second image of the atom and measure its final position. We repeat
the same experiment about 400 times with a single atom each time.

\begin{figure}[tb]
    \begin{center}
        \includegraphics[width=57mm]{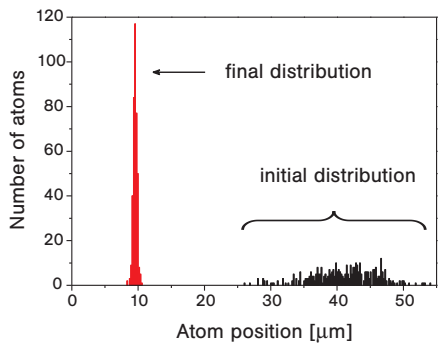}
    \end{center}
\caption{Absolute position control of single trapped atoms. The
histogram shows the accumulative data of about 400 experiments
carried out with one single atom at a time. Transfer of the atoms
from the MOT to the DT yields the broad distribution on the right
(standard deviation $5.0\ \mu$m). We transport the atoms to the
target position at $x_\mathrm{target}= 9.5\ \mu$m with a precision
of 300~nm rms (narrow distribution on the left).}
    \label{fig:PositionFeedback}
\end{figure}

Because the atoms are randomly loaded from the MOT into the DT,
the distribution of their initial positions, see
Fig.~\ref{fig:PositionFeedback}, has a standard deviation of
$5.0\,(\pm0.3)\,\mu$m, corresponding to the MOT radius. After the
transport, the width of the distribution of the final positions is
drastically reduced to
$\sigma_{\mathrm{control}}=300\,(\pm15)$~nm. This width is limited
by the errors in determining the final and initial position of the
atom, by the transportation error $\sigma_{\mathrm{transp}}$,
resulting from the discretization error of our digital
dual-frequency synthesizer which drives the optical conveyor belt,
and by the DT position drifts, $\sigma_{\mathrm{drift}}$, during
the typical time of 1.5~s between the two successive exposure
intervals. From the above DT phase measurement we find
$\sigma_{\mathrm{drift}}=140 \,(\pm 20)$~nm. Assuming that
    \begin{equation}\label{eq:transport}
        \sigma_{\mathrm{control}}= \sqrt{2\Delta x_\mathrm{stat}^2+
        \sigma_{\mathrm{drift}}^2+ \sigma_{\mathrm{transp}}^2},
    \end{equation}
we calculate that $\sigma_\mathrm{transp}=190\,(\pm 25)$~nm,
comparable to the statistical error.

In addition to statistical errors, the accuracy of the position
control is subject to systematic errors. The predominant
systematic error stems from the calibration of our length scale.
In the present case, a relative calibration error of $0.4$~\%
results in a $120$~nm shift of the final positions with respect to
the target position after a transport over $L\approx30\,\mu$m.
However, this error could be reduced by improving the accuracy of
the calibration.

Summarizing, we have realized a detection scheme for the absolute
and relative position of individual atoms stored in our standing
wave dipole trap, yielding sub-micrometer resolution. We have
shown that this scheme allows us to measure the exact number of
potential wells separating simultaneously trapped atoms in our
532~nm-period standing wave potential. We have furthermore used
our position detection scheme to transport an atom to a
predetermined position with a sub-optical wavelength accuracy.
These results represent an important step towards experiments in
which the relative or absolute position of single atoms has to be
controlled to a high degree. For example, we aim to use this
technique for a deterministic coupling of atoms to the mode of a
high-Q optical resonator in order to realize quantum logic
operations~\cite{Walther01, Blatt02, Schrader04, You03}.
Furthermore, knowing the exact number of potential wells
separating the atoms, we can now attempt to control this parameter
by placing atoms into specific potential wells of our standing
wave using additional optical tweezers. Finally, the demonstrated
high degree of control allows us to envision the implementation of
controlled cold collisions between optically resolved individual
atoms by means of spin-dependent transport \cite{Mandel03}.

We acknowledge valuable discussions with V.~I.~Balykin. This work
was supported by the Deutsche For\-schungs\-ge\-mein\-schaft and
the EC (IST/FET/QIPC project ``QGATES''). I.~D.~acknowledges
funding from INTAS. D.~S.~acknowledges funding by the Deutsche
Telekom Stiftung.


\begin{thebibliography}{99}

\bibitem{Binnig99}
 G. Binnig and H. Rohrer,
 Rev. Mod. Phys. {\bf 71}, S324 (1999)

\bibitem{Giessibl03}
 F. J. Giessibl,
 Rev. Mod. Phys. {\bf 75}, 949 (2003)

\bibitem{Suenaga00}
 K. Suenaga {\it et al.},
 Science {\bf 290}, 2280 (2000)

\bibitem{DiVincenzo95 and Ekert96}
 D. P. DiVincenzo,
 Science {\bf 270}, 255 (1995);
 A. Ekert and R. Jozsa,
 Rev. Mod. Phys. {\bf 68}, 733 (1996)

\bibitem{Walther01}
 G. R. Guth\"ohrlein {\it et al.},
 Nature {\bf 414}, 49 (2001)

\bibitem{Blatt02}
 A. B. Mundt {\it et al.},
 Phys. Rev. Lett. {\bf 89}, 103001 (2002)

\bibitem{Wineland03}
 D. Leibfried {\it et al.},
 Nature {\bf 422}, 412 (2003)

\bibitem{Thomas95}
 J. E. Thomas and L. J. Wang,
 Phys. Rep. {\bf 262}, 311 (1995)

\bibitem{Kuhr01}
 S. Kuhr {\it et al.},
 Science {\bf 293}, 278 (2001)

\bibitem{Bergamini04}
 S. Bergamini {\it et al.},
 J. Opt. Soc. Am. B {\bf 21}, 1889 (2004)

\bibitem{Schrader01}
 D. Schrader {\it et al.},
 Appl. Phys. B {\bf 73}, 819 (2001)

\bibitem{Miroshnychenko03}
 Y. Miroshnychenko {\it et al.},
 Optics Express {\bf 11}, 3498 (2003)

\bibitem{Alt02}
 W. Alt,
 Optik {\bf 113}, 142 (2002)

\bibitem{Mandel03}
 O. Mandel {\it et al.},
 Nature {\bf 425}, 937 (2003)

\bibitem{Scheunemann00}
 R. Scheunemann {\it et al.},
 Phys. Rev. A {\bf 62}, 051801(R) (2000)

\bibitem{Schrader04}
 D. Schrader  {\it et al.},
 Phys. Rev. Lett. {\bf 93}, 150501 (2004)

\bibitem{You03}
 L.~You, X.~X. Yi, and X.~H. Su,
 Phys. Rev. A {\bf 67}, 032308 (2003)

\end{thebibliography}
\end{document}